\documentclass[%
 aip,
 amsmath,amssymb,
 reprint,%
floatfix,
]{revtex4-2}

\usepackage{graphicx}
\usepackage{dcolumn}
\usepackage{bm}

\usepackage[utf8]{inputenc}
\usepackage[T1]{fontenc}
\usepackage{mathptmx}
\usepackage{etoolbox}

\usepackage{xcolor}

\usepackage{float}

\usepackage{mathtools} 
\usepackage{hyperref}

\usepackage{xcolor}
\usepackage[inline]{enumitem}

\newlength{\subcolumnwidth}

\newcommand{\nextsubcolumn}[1][]{%
  \cr\noalign{\hfill}
  \if\relax\detokenize{#1}\relax\else\hsize=#1\setlength{\subcolumnwidth}{\hsize}\fi
}

\makeatletter
\def\@email#1#2{%
 \endgroup
 \patchcmd{\titleblock@produce}
  {\frontmatter@RRAPformat}
  {\frontmatter@RRAPformat{\produce@RRAP{*#1\href{mailto:#2}{#2}}}\frontmatter@RRAPformat}
  {}{}
}%
\makeatletter
\def\@email#1#2{%
 \endgroup
 \patchcmd{\titleblock@produce}
  {\frontmatter@RRAPformat}
  {\frontmatter@RRAPformat{\produce@RRAP{*#1\href{mailto:#2}{#2}}}\frontmatter@RRAPformat}
  {}{}
}

\draft 

\begin{document}


\title{Inverse Design of Crystals and Quasicrystals in a  Non-Additive Binary  Mixture of Hard Disks} 



\author{Edwin A. Bedolla-Montiel}
 \affiliation{Soft Condensed Matter \& Biophysics, Debye Institute for Nanomaterials Science, Utrecht University, Princetonplein 1, 3584 CC Utrecht, Netherlands.}
\author{Jochem T. Lange}
\affiliation{Soft Condensed Matter \& Biophysics, Debye Institute for Nanomaterials Science, Utrecht University, Princetonplein 1, 3584 CC Utrecht, Netherlands.}%
\author{Alberto Pérez de Alba Ortíz}
  \affiliation{Computational Soft Matter Lab, Computational Chemistry Group and Computational Science Lab, van 't Hoff Institute for Molecular Science and Informatics Institute, University of Amsterdam, Science Park 904, 1098 XH Amsterdam, Netherlands.}
 \affiliation{Soft Condensed Matter \& Biophysics, Debye Institute for Nanomaterials Science, Utrecht University, Princetonplein 1, 3584 CC Utrecht, Netherlands.}
\author{Marjolein Dijkstra}
 \affiliation{Soft Condensed Matter \& Biophysics, Debye Institute for Nanomaterials Science, Utrecht University, Princetonplein 1, 3584 CC Utrecht, Netherlands.}
 \email{m.dijkstra@uu.nl, e.a.bedollamontiel@uu.nl}
\date{\today}

\begin{abstract}
The development of new materials typically involves a process of trial and error, guided by insights from past experimental and theoretical findings.
The inverse design approach for soft-matter systems has the potential to optimize specific physical parameters such as particle interactions, particle shape, or composition and packing fraction. This optimization aims to facilitate  the spontaneous formation of specific target structures through self-assembly. 
In this study, we expand upon a recently introduced  inverse design protocol for monodisperse systems to identify the required conditions and interactions for assembling crystal and quasicrystal phases within a binary mixture of two distinct species.
This method utilizes an evolutionary algorithm to identify the optimal state point and interaction parameters, enabling the self-assembly of the desired structure.
Additionally, we employ a convolutional neural network (CNN) that classifies different phases based on their diffraction patterns, serving as a fitness function for the desired structure.
Using our protocol, we successfully inverse design two-dimensional crystalline structures, including a hexagonal lattice, and a dodecagonal quasicrystal, within a  non-additive binary mixture of  hard disks.
Finally, we introduce a symmetry-based order parameter that leverages the encoded  symmetry within the diffraction pattern.
This order parameter circumvents the need for training a CNN, and is used as a fitness function to inverse design an octagonal quasicrystal.
\end{abstract}


\maketitle 

\section{\label{sec:intro} Introduction}
The self-assembly of colloidal particles  is a pivotal mechanism for fabricating nanostructured materials.
Colloidal systems, consisting of  particles ranging from nanometer to micrometer sizes  suspended in a fluid medium, inherently possess the ability to spontaneously organize themselves into structured arrangements due to interparticle  forces and thermodynamic conditions.
These systems show promise for potential applications due to their  photonic,~\cite{hynninenSelfassemblyRoutePhotonic2007,heColloidalDiamond2020} magnetic and  electronic properties.~\cite{deguchiQuantumCriticalState2012}
However, the structure of the assembly of such materials depends on the building blocks, namely the interaction between the species composing the system, and the thermodynamic conditions, such as  temperature, pressure, density, or composition.
Understanding the relationship between these building blocks, thermodynamic state, and  self-assembled structures is crucial for leveraging  them for materials design. This relationship is fundamental, as the physical properties of each material are intrinsically intertwined with its  structure.

In the forward design approach, a specific colloidal particle system is chosen as the foundational building blocks for a material with desired properties.
Subsequently,  the interaction parameters and  thermodynamic conditions are systematically changed until achieving the desired material.
The forward design approach can quickly become unfeasible, because the number of possible building block combinations and  conditions needed to assemble the required structures is intractably large.

Recently, there has been a growing  interest in developing frameworks for the inverse design of self-assembled  structures and materials.~\cite{dijkstraPredictiveModellingMachine2021}
The inverse design approach directly determines the parameters and thermodynamic conditions necessary to attain  a target structure with specific properties.~\cite{Sherman_2020}
Inverse design protocols have been successfully applied to the search of both crystalline~\cite{rechtsmanOptimizedInteractionsTargeted2005,lieuFormationFluctuationTwodimensional2022} and quasicrystalline structures.~\cite{zuFormingQuasicrystalsMonodisperse2017,maInverseDesignSelfassembling2019,coliInverseDesignSoft2021,lieuFormationFluctuationTwodimensional2022,lieuInverseDesignTwodimensional2022a,yoshinagaBayesianModelingPattern2022,noya2021design}
One category of methods involves adjusting intermolecular interactions to  target the radial distribution function of the desired phase. The objective is to precisely  match the two-body structural correlations using a maximum entropy optimization scheme.~\cite{b.jadrichEquilibriumClusterFluids2015,pinerosInverseDesignMulticomponent2018}

Recently, machine learning methods have been used for  the analysis, identification, and formation of soft-matter systems.~\cite{dijkstraPredictiveModellingMachine2021}
Coli \emph{et al.} developed an inverse design protocol~\cite{coliInverseDesignSoft2021} employing supervised machine learning techniques.
The authors used a convolutional neural network (CNN) as a classifier  to  differentiate structures based on their diffraction patterns.
Subsequently, target structures are achieved by modifying the interaction parameters and thermodynamic state points through an evolutionary strategy optimization algorithm, which uses a CNN-based fitness.
This exploration helps identifying the thermodynamic conditions and interactions necessary for stabilizing  the desired structures.
The methodology proposed by  Coli \emph{et al.} facilitates the  self-assembly of quasicrystals, as the symmetry of the structure is encoded within the diffraction patterns.
Recently, Lieu and Yoshinaga proposed an alternative framework in which reinforcement learning is used along with patchy particles to promote the self-assembly of dodecagonal quasicrystals.~\cite{lieuDynamicControlSelfassembly2023}
Using an optimization method tailored for reinforcement learning, they  automatically tuned the cooling schedule such that a critical temperature is identified at which the quasicrystal is formed.

In this work, we expand upon the protocol of Coli \emph{et al.} to encompass non-additive binary mixtures of hard disks, a system more akin to the quasicrystals observed in the experiments on monolayers of inorganic nanoparticles.~\cite{talapinQuasicrystallineOrderSelfassembled2009}
The phase behavior of a similar system, \textit{i.e.} binary mixtures of hard spheres, has been explored by computer simulations.~\cite{fayenSelfassemblyDodecagonalOctagonal2023a,fayenInfinitepressurePhaseDiagram2020a}
We employ a CNN to classify  different structures using the diffraction patterns of one of the species within the mixture.
Subsequently, we use an evolutionary strategy algorithm to optimize the thermodynamic conditions and interaction parameters, facilitating the self-assembly of binary crystals and quasicrystals.
However, this protocol encounters severe challenges when the training data for the CNN lacks  diffraction patterns of the target structure, hindering its effectiveness in inverse designing the desired system.
To address this limitation, we introduce a novel symmetry-based order parameter.
This order parameter eliminates the necessity for training  a CNN and bypasses the requirement for diffraction patterns of the  structures to be inverse-designed. 
We observe that this order parameter, which measures the order of symmetry of the diffraction pattern, can successfully be used as a fitness function in the evolutionary strategy algorithm to optimize the physical parameters for the self-assembly of the desired structures. 

This paper is organized as follows.
In Sec.~\ref{sec:model}, we introduce our model for the binary mixture of hard disks. The inverse design protocol is presented in Sec.~\ref{sec:inverse}.
This protocol comprises simulations for sampling, both a CNN and a symmetry-based order parameter for  fitness evaluation, and an evolutionary strategy for the optimization of the parameters. 
In Sec.~\ref{sec:results}, we present the main results, illustrating the self-assembly of  crystalline and quasicrystalline structures via the CNN approach.
Following this, we employ the symmetry-based order parameter to successfully inverse design an octagonal quasicrystal.
We present our conclusions in Sec.~\ref{sec:conclusions}.

\section{\label{sec:model} Model}
In our inverse-design approach for self-assembly of binary mixtures, we aim to produce trajectories that start in a disordered, low-density phase and finish in the target crystalline or quasicrystalline phase.
Producing such trajectories requires time propagation of particle motions, which we compute via molecular dynamics (MD) simulations.~\cite{frenkel2002understanding} 
We consider a two-dimensional non-additive binary mixture of  hard disks of two different sizes.  
The large ($L$) species have a diameter $\sigma_L$ and the small ($S$) species a diameter $\sigma_S$.
To perform MD simulations of hard-disk systems, it is necessary to employ a continuous interaction potential.
Using the Extended Law of Corresponding States as formulated by Noro and Frenkel,~\cite{noro2000} and assuming its applicability extends to repulsive potentials, B\'{a}ez \emph{et al.}~\cite{baezUsingSecondVirial2018} argued that by using a re-parametrization of the intermolecular potential proposed by Jover \emph{et al.},~\cite{jover2012} it becomes feasible to map the hard-core interaction onto a continuous potential.
This approach  yields accurate results, particularly when the continuous potential precisely reproduces the second virial coefficient of the true hard-particle potential.
Such a re-parametrization of the continuous hard-core potential  $u_{\alpha\beta}(r)$ between species $\alpha = L,S$ and species $\beta=L,S$ reads
\begin{eqnarray}
    u_{\alpha\beta}(r) &=& \nonumber \\
    & & \hspace{-8mm}
    \begin{cases}
        A \, \epsilon \left[ {\left(\frac{\sigma_{\alpha\beta}}{r}\right)}^{\lambda} -
        {\left(\frac{\sigma_{\alpha\beta}}{r}\right)}^{\lambda - 1} \right] + \epsilon & ,
        r < \sigma_{\alpha\beta} \, B \\
        \hspace{2cm} 0 & , r \geq \sigma_{\alpha\beta} \, B
    \end{cases}
    \, ,
    \label{eq:cont-hs}
\end{eqnarray}
with
\begin{equation}
    A = \lambda {\left(\frac{\lambda}{\lambda -1}\right)}^{\lambda - 1} \, ;
    \quad
    B = \left(\frac{\lambda}{\lambda -1}\right) \, ,
    \label{eq:ab-params}
\end{equation}
where $r$ denotes the center-of-mass distance between the two particles, $\sigma_{\alpha \beta}$ represents the hard-core diameter between species $\alpha$ and $\beta$, and $\epsilon$ denotes the effective strength of the interaction.
We set the hard-core diameters $\sigma_{LL}=\sigma_L$, $\sigma_{SS}=\sigma_S$, and introduce a non-additivity parameter $\Delta$ for the contact distance between the large and small species
\begin{equation}
    \sigma_{LS} = \frac{\sigma_S + \sigma_L}{2} (1 - \Delta) \, .
    \label{eq:nonadd}
\end{equation}
A non-linear equation in terms of the reduced temperature $T^{*} = k_{B} T / \epsilon$ with $k_{B}$ Boltzmann's constant must be solved in order to determine the  temperature at which the difference between the second virial coefficient of both interaction potentials becomes zero.

In this study, the parameter $\lambda$ is held constant at $\lambda=50$ in Eq.~\eqref{eq:cont-hs}, as well as the temperature, which remains constant across  all simulations.
For the two-dimensional systems studied here, the temperature is set to $T^{*}=1.4671$ following the results of Ba\'{e}z \emph{et al.}~\cite{baezUsingSecondVirial2018} for a two-dimensional system.
By fixing $\lambda$, the values of $A$ and $B$ are also fixed in Eq.~\eqref{eq:ab-params}.

The thermodynamic state of our binary mixture is defined by four design parameters:

\begin{itemize}
    \item the size ratio $q=\sigma_{S} / \sigma_{L}$,
    \item the small species composition $x_{S} = N_{S} / N$ with $N_{S}$ the number of particles for the small species,
    \item the non-additivity parameter $\Delta$,
    \item and the packing fraction $\eta = (N_{S} \, \sigma_{S}^{2} + N_{L} \, \sigma_{L}^{2}) \, \pi / 4A$ with $A$ the area defined by the simulation box.
\end{itemize}

\section{\label{sec:inverse} Inverse Design Protocol}
Our aim is to optimize the interaction parameters as well as the thermodynamic state point to favor the self-assembly of specific target phases in a non-additive binary hard-disk mixture. 
In this work, we build upon the inverse design protocol introduced in Ref.~\citenum{coliInverseDesignSoft2021}.
This approach combines  the  covariance matrix adaptation evolution strategy (CMA-ES) to sample and optimize  the set of design parameters, using a CNN to evaluate  the fitness of each sample. 
The method consists of three steps. 
In the first step of the inverse design process, a fixed number of parameter sets are drawn from a multivariate Gaussian distribution.
The dimension of the multivariate Gaussian distribution is set by the number of free design parameters that need to be tuned.
For each set of parameters, also called sample, we perform MD simulations of the system from which equilibrated configurations will be collected to compute  diffraction patterns.
In the second step, we evaluate the fitness of the samples  by classifying their diffraction patterns using the CNN or the symmetry-based order parameter. 
Samples more likely to be classified  as the target phase receive a higher fitness value. 
In the final step, we update the mean and covariance matrix of the Gaussian distribution  to move towards regions within the parameter space where the fittest samples are located.
Below, we describe the three steps in more detail. 

\subsection{Simulations}
We perform MD simulations of the non-additive binary hard-disk mixture in a square simulation box with periodic boundary conditions in all directions using the LAMMPS simulation code version 28 Mar 2023.~\cite{LAMMPS,Intveld08,Monti2022,Shire2020}
Each system consists of $N = N_{S} + N_{L}=512$  particles.
We initialize the system by randomly placing the particles within the simulation box at an initial packing fraction of $\eta=0.3$.
Subsequently, the simulation box is linearly compressed to a higher target packing fraction, determined by CMA-ES, over a duration of $10^{8} \, \tau$, where $\tau=\sigma_{L}\sqrt{m / \epsilon}$ denotes the simulation time unit, and $m$ represents the mass of a particle, which is kept identical for both species.
This box deformation scheduling is used to reach the higher packing fractions where the crystals and quasicrystals can be self-assembled.
A time step of $\tau = 0.001$ is employed  for all simulations.
Upon reaching the target packing fraction, an equilibration phase follows wherein the positions of the particles are propagated for at least $10^{7} \, \tau$  to reach equilibrium.
However, for systems exhibiting quasicrystal phases, at least $10^{8} \, \tau$ was required for equilibration, which we determine by monitoring the time evolution of the energy of the system and the radial distribution function.
Longer runs are needed in order to reduce the number of defects in the system.~\cite{doteraMosaicTwolengthscaleQuasicrystals2014a,schoberthMolecularDynamicsStudy2016}
Subsequently, a production phase of at least $10^{7} \, \tau$ is performed to collect a large number of independent configurations of the system. 
Configurations were collected every $10^{4} \, \tau$, resulting in  a total of $10^{3}$ configurations.
The various thermodynamic state points are explored using simulations in the canonical ($NVT$) ensemble with canonical thermostatting,  employing the method introduced by  Bussi \emph{et al}.~\cite{bussi2007}
The thermostat relaxation time is set to  100 $\tau$.

\subsection{Convolutional neural network as classifier}

\begin{figure*}
    \includegraphics[width=\textwidth]{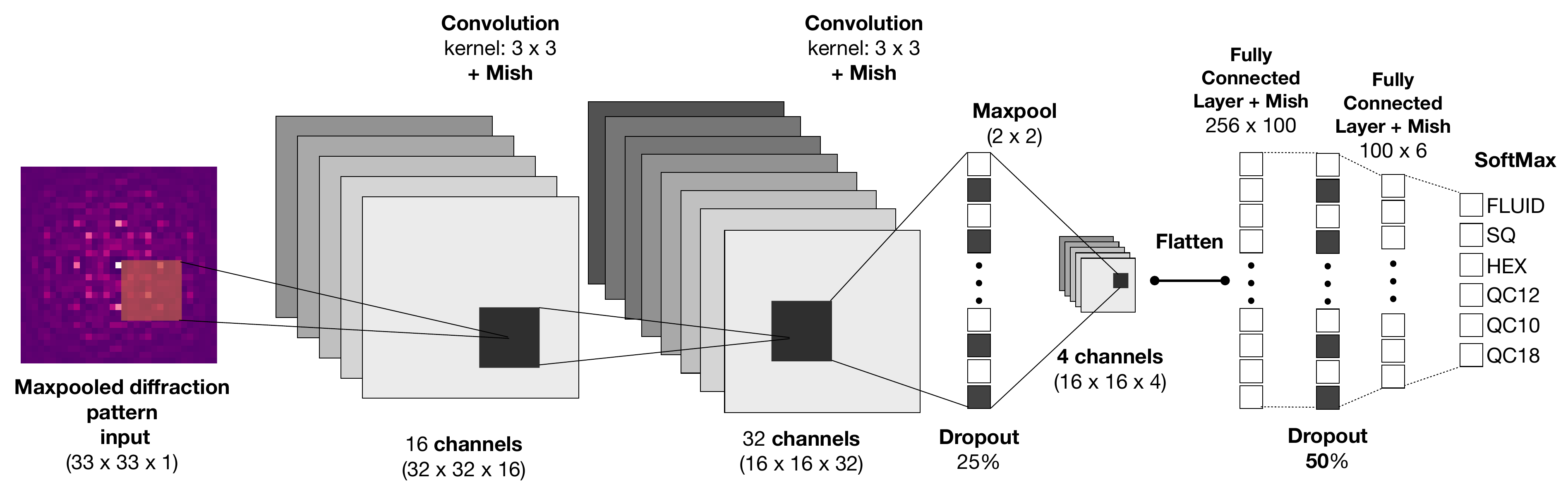}
    \caption{\textbf{Schematic representation of the  CNN used as a classifier in this work}. The network consists of two convolutional layers for feature extraction, followed by a MaxPooling layer to reduce resolution and two fully connected layers before a SoftMax function for final classification. All details about kernels, layer size, and activation functions are also shown. Each layer uses a non-linear \textbf{Mish} activation function. Note that between the fully connected layers there is a dropout layer to avoid overfitting.}
    \label{fig:schematic}
\end{figure*}

The inverse design protocol used in Ref.~\citenum{coliInverseDesignSoft2021} employs a CNN as a classifier.
The output of the CNN provides a probability indicating how closely  a given configuration resembles the target phase, effectively serving as a fitness function. 
Subsequently, a derivative-free optimization method is employed to optimize the fitness of the samples. 
The CNN takes as input the diffraction patterns extracted from the configurations gathered during the simulations. 
The aim of employing a CNN is to classify the various phases within a  non-additive binary hard-disk mixture based on their respective diffraction patterns.
We anticipate this two-dimensional mixture to stabilize various phases, including, for example, an isotropic fluid phase (FLUID), hexagonal (HEX) and square (SQ) crystal phases, and quasicrystal phases such as the decagonal (QC10), dodecagonal (QC12), and octadecagonal (QC18) quasicrystal phases.  

To train the CNN,~\cite{coliInverseDesignSoft2021} we need diffraction patterns corresponding to these phases.
As the aim of this paper is to inverse design these phases within  a binary mixture of hard disks, these diffraction patterns are currently unavailable.
However,  data is available from a single-component system of particles interacting with a hard-core square-shoulder potential.~\cite{coliInverseDesignSoft2021} 
This monodisperse system exhibits crystalline and quasicrystalline phases sharing  similar translational and rotational symmetries as the anticipated phases in the binary mixture.~\cite{doteraMosaicTwolengthscaleQuasicrystals2014a,pattabhiramanPhaseBehaviourQuasicrystal2017,coliInverseDesignSoft2021} In this study, we investigate the possibility of training the CNN using diffraction patterns from a monodisperse two-dimensional system interacting via a hard-core square-shoulder potential for inverse designing these phases in a binary mixture.~\cite{doteraMosaicTwolengthscaleQuasicrystals2014a,coliInverseDesignSoft2021}

To generate training data, Monte Carlo simulations of the two-dimensional hard-core square-shoulder model were performed for a total of $5 \times 10^5$ Monte Carlo sweeps, where  each sweep represents an attempt to randomly displace all particles in the system.
The equilibration phase is followed by  $1 \times 10^6$ sweeps, during which we save a configuration every $10^3$ sweeps, resulting in $10^3$ independent configurations.
This is repeated for 10 different state points for each of the six considered phases, FLUID, HEX, SQ, QC12, QC10, and QC18.
To ensure identification of phases regardless of their orientation, every training configuration is rotated by a random angle before evaluating its diffraction pattern following the procedure outlined in Ref.~\citenum{coliInverseDesignSoft2021}.
The diffraction patterns for these one-component systems are obtained by computing the two-dimensional structure factor,
\begin{equation}
    S({\mathbf{k}}) = \frac{1}{N} {\left\vert \sum_{j=1}^{N} e^{i \, {\mathbf{k}} \cdot {\mathbf{r}}_{j}} \right\vert}^{2} \, ,
    \label{eq:stucture-factor}
\end{equation}
where ${\mathbf{k}} = 2 \pi (n_x,n_y)/L$ represents the wavevector with $n_x$ and $n_y$ denoting two integers, $L$ the box length, and ${\mathbf{r}}_{j}$ the center-of-mass position of particle $j$. We determine the diffraction pattern for each training configuration on a $150 \times 150$ grid, followed by a size reduction using a max pooling filter to achieve a final size of $33 \times 33$.
The primary goal of max pooling is to reduce the dimensions of the data while preserving important input features.
Consequently, the max pooling step reduces the computational time and  memory during training.

To classify and evaluate structures in our binary mixture, we consider only the large species in computing the diffraction patterns.
The CNN used in this work is comprised of two convolutional layers for feature extraction and two fully connected layers for classification as depicted schematically in Fig.~\ref{fig:schematic}.
Each convolutional layer performs two operations: feature extraction through convolutional filters and a non-linear transformation using the \emph{Mish} activation function, defined as $f(x)=x \, \tanh{\left[\log{\left(1 + e^{x}\right)}\right]}$.~\cite{misra2020mish}
The first convolutional layer has one input channel, corresponding to the maxpooled diffraction pattern, and sixteen output channels, which are the extracted features.
The convolutional layer uses kernels of size $3 \times 3$, with a stride of $s=1$ and a padding of $p=1$.
The second convolutional layer takes as input the sixteen channels and outputs thirty-two channels in order to increase the features extracted from the images.
The kernels in this layer have the same size, padding and stride as the first convolutional layer.
We use a downsampling step after the channels through a layer with a $2 \times 2$ max pooling filter with a stride of $s=2$.
To enhance the robustness of the neural network, a dropout probability of $P=0.25$ is introduced following the downsampling of the second convolutional layer. 
Dropout is a regularization technique that selectively eliminates feature detectors with a probability $P$, while the remaining weights are trained as usual.~\cite{NIPS2013_71f6278d}
The dropout layer is subsequently stacked, flattened, and prepared as input for classification, where two fully connected layers are used to classify the features extracted from the convolutional layers.
The first fully connected layer consists of 256 units, and the second layer consists of 100 units.
The \emph{Mish} activation function is applied to all units in the layers.
To further enhance robustness and introduce regularization, a dropout probability of $P=0.5$ is applied between the two fully connected layers.
The output layer contains six units to align with the available training classes, followed by a SoftMax function to determine classification probabilities.

We train the CNN  by minimizing the cross-entropy loss while implementing $L_{2}$ regularization  as weight decay of $10^{-5}$.~\cite{Goodfellow-et-al-2016} This regularization of weights enhances robustness of the network, making it less prone to overfitting.
The Adam optimizer~\cite{kingma2017adam} is used with a learning rate of $10^{-3}$.
To prevent overfitting, early-stopping is employed.
If the loss remains unchanged between consecutive epochs, the training is halted and the optimization is assumed to have reached convergence within a maximum of fifty epochs.
The dataset comprises a total of $48000$ samples, divided into $8000$ samples per class,  split in an $80-20\%$ ratio for training and test sets.
The test set, containing samples not used during training, is employed to evaluate the performance of the classifier, which reaches a final classification accuracy in the test set of at least $99.9 \%$.
The CNN implementation and training protocol are carried out using PyTorch and PyTorch Lightning.~\cite{Paszke_PyTorch_An_Imperative_2019,Falcon_PyTorch_Lightning_2019}

\subsection{\label{sec:DSASS} Symmetry-Based Order Parameter}
\begin{figure}
    \includegraphics[width=\linewidth]{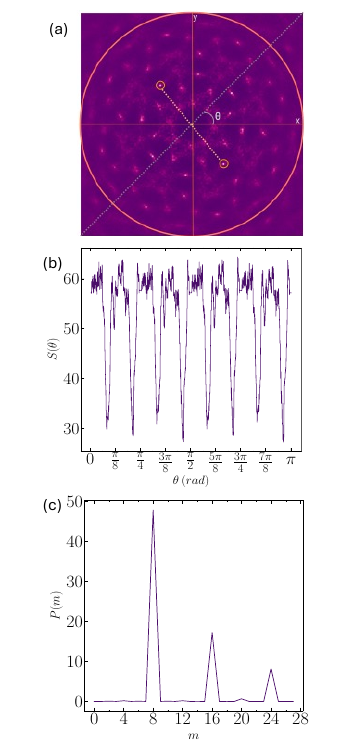}
    \caption[score]{\textbf{Symmetry-based order parameter.}
    \begin{enumerate*}[label=(\alph*)]
        \item A grey line represents the reflection axis, forming an angle $\theta$ with the $x$-axis. Intensities of points from one half of the circle are subtracted point-wise from the intensity of their reflection points across the reflection axis on the other half of the circle. An example of intensities are depicted by small circles and connected by a dotted line in the image. The absolute sum of all intensity differences quantifies the degree of inverse reflectivity $S(\theta)$.
        \item $S(\theta)$ as a function of $\theta$.
        \item The power spectral density $P(m)$ of $S(\theta)$ as a function of $m$, with $m$ denoting the $m$-fold rotational symmetry.
     \end{enumerate*}}
    \label{fig:sscore}
\end{figure}

It is important to note that training the CNN relies on having the diffraction patterns for the target structures.
However, when we aim to reverse engineer a structure without access to its diffraction pattern, this inverse design method will not succeed.
To overcome this challenge, we introduce a symmetry-based order parameter leveraging the unique rotational symmetry within diffraction patterns, eliminating the need for training a CNN, and serving as a fitness function for inverse design protocols.
To analyze the symmetry present in a diffraction pattern, we ascertain the count of detectable symmetry axes within the aforementioned pattern.
More specifically, we identify the lines of symmetry that cut the diffraction pattern into two identical mirror images.
To discern lines of symmetry, we posit a metric to quantify the degree of inverse reflectivity exhibited by a given axis.
Mathematically, this reads 
\begin{equation}
    S(\theta) = \sum_{\vec{r} \in A} \left\vert I \left( Q(\theta) \cdot \mathbf{r} \right) - I(\mathbf{r}) \right\vert \, ,
    \label{eq:symmetry-score}
\end{equation}
where $A = \{(x, y) \colon x^2 + y^2 \leq R^2  \}$ represents  all points $\mathbf{r} = (x, y)$ lying within a circle of radius $R$ centered around the origin of the diffraction pattern.
The angle $\theta$ denotes the angle formed between the reflection axis and the $x$-axis as illustrated in Fig.~\ref{fig:sscore}(a).
The peak intensity within the diffraction pattern at  location $\mathbf{r}$ is defined as $I(\mathbf{r})$.
Finally, $Q(\theta)$ represents the reflection matrix across the reflection axis at an angle $\theta$ with the $x$-axis
\begin{equation}
    Q(\theta) = \begin{pmatrix}
    \cos{(2 \theta)} & \sin{(2 \theta)} \\
    \sin{(2 \theta)}  & -\cos{(2 \theta)} 
    \end{pmatrix} \, .
    \label{eq:reflection}
\end{equation}
By inverse reflectivity we mean that a high value of $S(\theta)$ indicates a lack of reflection symmetry for that particular angle $\theta$, i.e. there is no mirror symmetry between the two opposite sections of the diffraction pattern.
Conversely, a low value of $S(\theta)$ suggests high reflection symmetry for that specific angle $\theta$.

Applying the floor operation to  the expression $Q(\theta) \cdot \mathbf{r}$ in Eq.~\eqref{eq:symmetry-score} is essential to ensure the use of  real pixel locations  when dealing with the  raw data from diffraction patterns.
Fig.~\ref{fig:sscore}(a) provides a schematic representation of  this method,  illustrating the circle of radius $R$ in the diffraction pattern.
We note that as the method operates with real information from diffraction pattern images, it can readily be extended to experimental setups.

We calculate $S(\theta)$ for a uniform grid of $M$ angles in the range of $\theta \in [0, \pi)$. 
We show $S(\theta)$ as a function of $\theta$ in Fig.~\ref{fig:sscore}(b), where we find an oscillatory pattern in $S(\theta)$ as a function of $\theta$.
The oscillation frequency of $S(\theta)$ correlates with the number of reflection axes of the diffraction pattern.
To evaluate the number of symmetry axes, we compute the power spectral density  $P(m)$.
First, we subtract the mean from $S(\theta)$ in order to normalize the symmetry score across different samples.
Subsequently, the power spectral density $P(m)$ is computed according to the equation 
\begin{equation}
    P(m) = \frac{1}{M} {\left\vert \sum_{k=0}^{M - 1} S(\theta_{k}) \, e^{-i 2 \pi m k \Delta \theta} \right\vert}^{2} \, ,
    \label{eq:psd}
\end{equation}
where  $\theta_{k} = \theta_{0} + k \Delta \theta$ represents the discrete angles used to compute the discrete symmetry scores $S(\theta_{k})$, with $\theta_{0} = 0$ and $k \in \{ 0, 1, 2, \dots, M - 1 \}$, $i$ denotes the imaginary unit, $m \in \{ 0, 1, 2, \dots M - 1\}$ represents the $m$-fold rotational symmetry, and ${\lvert \cdot \rvert}^{2}$ denotes the norm of the complex numbers.
Essentially, the power spectral density $P(m)$ decomposes the signal into its Fourier components such that the underlying rotational symmetries are easier to detect.
A representative plot of the power spectral density $P(m)$ is shown in Fig.~\ref{fig:sscore}(c), where the high peaks correspond to the rotational symmetries obtained from the signal decomposition of $S(\theta)$.
For an $m$-fold rotationally symmetric diffraction pattern, we expect to find a high value at the $m$ value of the power spectral density $P(m)$.
For instance, a QC8 exhibits a 8-fold rotational symmetry in its diffraction pattern, and the power spectral density $P(m)$  exhibits the 8-fold symmetry by having the largest value at $m=8$, as shown in Fig.~\ref{fig:sscore}(c).
It is essential to note that signal aliasing might occur, leading to high values at frequencies that are multiples of the sampling frequency, like the 16-fold and 24-fold symmetry for an 8-fold diffraction pattern (see the small peaks in Fig.~\ref{fig:sscore}(c)).~\cite{stoica2005spectral}

We can now define a symmetry-based order parameter $f_{m}$,  serving as a fitness function
\begin{equation}
    f_{m} = P(m) - \sum_{\substack{i=0 \, , \\ i \neq m}}^{M - 1} P(i) \, ,
    \label{eq:fitnessfunction}
\end{equation}
where we anticipate a high value for $P(m)$ when the diffraction pattern displays $m$-fold symmetry.
By subtracting all rotational symmetries that are not equal to $m$ (the second term in Eq.~\eqref{eq:fitnessfunction}), we penalize symmetries that might induce aliasing effects of the $m$-fold symmetry.
Subsequently, we can optimize $f_{m}$ defined in Eq.~\eqref{eq:fitnessfunction} using an evolution strategy algorithm, similar to the fitness value as determined by the CNN.
In this way, we promote the self-assembly of structures with $m$-fold symmetry.
It is important to note here that $f_{m}$ is not bounded from above, as opposed to the CNN, so the highest value of the peak is unknown beforehand.

\subsection{Covariance Matrix Adaptation Evolution Strategy as Optimizer}
In this study, we employ the  covariance matrix adaptation evolution strategy (CMA-ES) for optimizing the fitness function, which provides a measure of how close the system is to the target phase.
Here we present a short summary of the method. For further explanations, we direct the reader to the references cited in the following.
The CMA-ES represents a stochastic zero-order optimization technique designed to optimize a real-valued, non-convex, and nonlinear function, operating  without requiring gradient information of the function.~\cite{hansenCMAEvolutionStrategy2016}
CMA-ES is an iterative algorithm which samples from a multivariate Gaussian distribution and updates the mean and the covariance matrix of the Gaussian distribution at each iteration. This process continues until a convergence criterion is met. In each iteration, often referred to as a \emph{generation}, the algorithm draws $n$ samples from a $d$-dimensional multivariate Gaussian distribution, where $d$ represents the number of design parameters we wish to optimize. 
The fitness function is then evaluated for these $n$ samples, and the outcomes are arranged in descending order. Subsequently, the top  $k$ samples  from this particular generation are chosen as the best candidates, forming the elements of the set $\bm{X}$.
In each generation, the mean vector $\boldsymbol{\mu} \in \mathbb{R}^{d}$ and covariance matrix $\boldsymbol{C} \in \mathbb{R}^{d \times d}$ are updated  to adjust the Gaussian distribution  to reach the optimum.

Various proposals for updating these parameters have been introduced,~\cite{varelasComparative2018} but here we focus on the separable CMA-ES (sepCMA-ES).~\cite{rosSimpleModificationCMAES2008a}
The sepCMA-ES version constrains the covariance matrix to remain diagonal, enabling linear space and time complexity. The sepCMA-ES version outperforms the traditional CMA-ES algorithm in many optimization problems.~\cite{hansenBenchmarkingBIpopulationCMAES2009} 

We start from an initial mean vector $\boldsymbol{\mu}$ sampled uniformly within the feasible parameter space, and an initial covariance matrix $\boldsymbol{C} = \boldsymbol{I}$, with $\boldsymbol{I}$ the identity matrix with shape $d \times d$.
The sepCMA-ES generates a new candidate solution using the following equations 
\begin{align}
    \boldsymbol{z}_{i} &\sim \mathcal{N}(0, \boldsymbol{I}) \quad \text{for} \quad i = 1, \dots , n \nonumber \\
    \boldsymbol{x}_{i} &= \boldsymbol{\mu} + \sigma \boldsymbol{B} \boldsymbol{D} \boldsymbol{z}_{i} \nonumber \\
    \boldsymbol{\mu} &= \sum_{i=1}^{k} w_{i} \boldsymbol{x}_{i:k} \nonumber \\
    \boldsymbol{z} &= \sum_{i=1}^{k} w_{i} \boldsymbol{z}_{i:k}
    \label{eq:cmaes-mean}
\end{align}
where $\boldsymbol{x}_{i:k}$ denotes the $i$-th best individual out of the $k$ samples.
The elements of $\boldsymbol{z}_{i}$ are standard normal random variables. 
The matrix $\boldsymbol{B}$ has the orthogonal eigenvectors of $\boldsymbol{C}$ as columns, whereas
the matrix $\boldsymbol{D}$ has the eigenvalue square roots as diagonal elements.
The parameter $\sigma$ is the initial standard deviation used at the start of the optimization method.
The distribution of weights is defined as 
\begin{equation}
    w_i = \frac{\log{(k+1) - \log{(i)}}}{\sum_{j=1}^{R} \log{(k+1) - \log{(j)}}} \, ,
    \label{eq:weights}
\end{equation}
where $i$ is the rank index of sample $x$, with $i=1$ indicating  the configuration with the highest $f$ value.
These equations update the candidate solution as well as the mean vector used to sample the multivariate normal distribution.

We now turn our attention to updating the covariance matrix, using the following equations
\begin{align}
    p_{\sigma} &= (1 - c_{\sigma}) p_{\sigma} + \sqrt{\mu_{W} c_{\sigma} (2 - c_{\sigma})}
    \boldsymbol{Bz} \nonumber \\
    p_{c} &= (1 - c_{c}) p_{c} + H_{\sigma} \sqrt{\mu_{W} c_{c} (2 - c_{c})}
    \boldsymbol{BDz} \nonumber \\
    \boldsymbol{C} &= (1 - c_{cov}) \boldsymbol{C} + \frac{1}{\mu_{cov}} c_{cov} p_{c} p_{c}^{\boldsymbol{T}} \nonumber \\ 
    &+ c_{cov} \left( 1 - \frac{1}{\mu_{cov}} \right) \sum_{i=1}^{k} w_{i} \boldsymbol{BDz}_{i:k} ({\boldsymbol{BDz}_{i:k}})^{\boldsymbol{T}} \, ,
    \label{eq:cmaes-cov}
\end{align}
where the Heaviside function $H_{\sigma} = 1$ when the condition $\frac{\lVert p_{\sigma} \rVert}{\sqrt{1 - {(1 - c_{\sigma})}^{2g}}} < \left(1.4 + \frac{2}{d+1}\right) E\left(\left\lVert \mathcal{N}(0, \boldsymbol{I}) \right\rVert \right)$ is met.
This is the condition that happens most often during the evolution path, but the other case $H_{\sigma} = 0$ can happen
as well, which indicates that the evolution path is stalled.
The $d$-dimensional vectors $p_{\sigma}$ control the amplitude of the covariance matrix, while the $d$-dimensional vector $p_{c}$ manipulates the directionality.
Additionally, $\langle \Vert \mathcal{N}(0, \boldsymbol{I}) \Vert \rangle$ represents the average length of a vector sampled from a standard multivariate normal distribution.
This quantity is used for step-size control.
All the other parameters are free parameters which are constant throughout the optimization procedure, and the values chosen are explained below.

Finally, we update the step size $\sigma$ along with the covariance matrix and its decomposition, using the following equations
\begin{align}
    \sigma &= \sigma \exp{\left( \frac{c_{\sigma}}{d_{\sigma}} \left( \frac{\lVert p_{\sigma} \rVert}{E\left(\left\lVert \mathcal{N}(0, \boldsymbol{I}) \right\rVert \right)} - 1 \right) \right)} \nonumber \\
    \boldsymbol{D}^{2} &= \text{diag}(\boldsymbol{C}) \, ,
    \label{eq:cmaes-sigma}
\end{align}
where $\text{diag}(\boldsymbol{C})$ is a diagonal matrix with the same diagonal elements as $\boldsymbol{C}$.

In this work, we use $n=24$ and $k=12$ for all our results.
The value of the initial standard deviation changes for each optimization problem, since it depends on the bounds of the $d$-dimensional search space ${[a, b]}^{d}$.
In all results presented here, $\sigma = 0.3 (b-a)$ where $a$ and $b$ might be different depending on the phase to inverse design.
The initial values for $p_{\sigma}$ and $p_{c}$ in Eq.~\eqref{eq:cmaes-cov} are null vectors.
All free parameters $c_{i}$, $\mu_{W}$ and $\mu_{cov}$ in Eq.~\eqref{eq:cmaes-cov} and  Eq.~\eqref{eq:cmaes-sigma} are set to the default values as outlined in the original work~\cite{rosSimpleModificationCMAES2008a} and in the Python implementation by Nomura and Shibata~\cite{nomura2024cmaes}
that we use in this work.
Finally, the stopping criterion of the optimization scheme is determined by setting the number of generations to fifty generations.
This is a common choice for stopping the optimization scheme in numerical optimization experiments, where the number of function evaluations is fixed.~\cite{rios2013derivative}

\section{\label{sec:results} Results}
\subsection{Inverse design of a hexagonal lattice}
\begin{figure*}
\includegraphics[width=\textwidth]{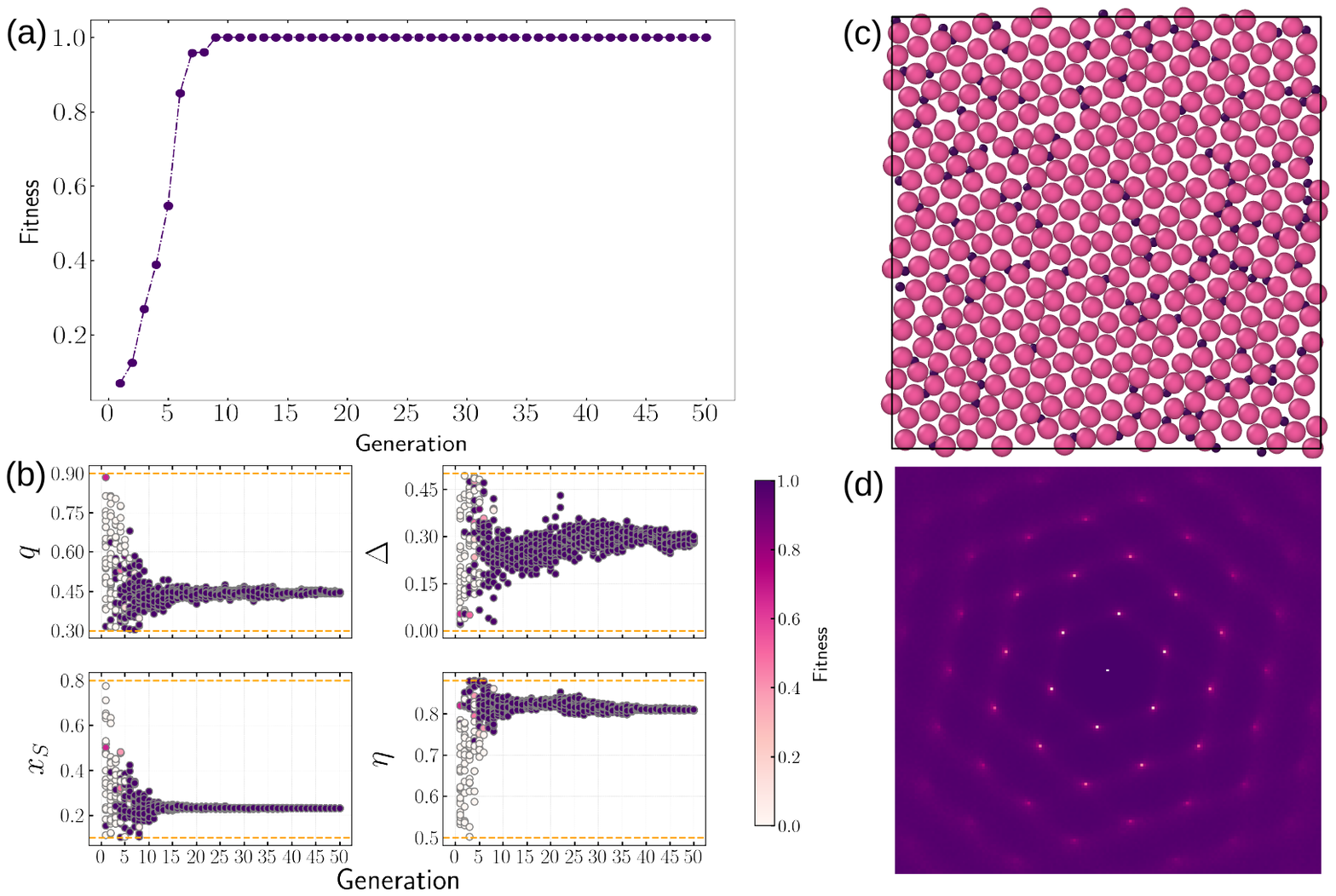}
    \caption[hex]{\textbf{Inverse design of the hexagonal lattice in a non-additive hard-disk mixture using the CNN as order parameter.}
        \begin{enumerate*}[label=(\alph*)]
         \item Evolution of the mean fitness during the inverse design protocol.
         \item Evolution of the design parameters, size ratio $q=\sigma_S/\sigma_L$, non-additivity parameter $\Delta$, small species composition $x_S$, and packing fraction $\eta$, for the inverse design of the hexagonal lattice. Each dot represents a sample for a given generation. The dotted lines are the bounds imposed for the optimization problem.
        \item Exemplary configuration snapshot of a hexagonal lattice obtained during the last generation. \label{conf-hex}
        \item Diffraction pattern of the configuration in~\ref{conf-hex}.
         \end{enumerate*}}
    \label{fig:fitness-hex}
\end{figure*}

We start our investigation by reverse-engineering the hexagonal crystal (HEX) phase in a non-additive binary mixture of hard disks.  
Our goal is to identify  the optimal set of design parameters that facilitates the self-assembly of a hexagonal lattice within this mixture.
To accomplish this, we use the size ratio $q$, the small species composition $x_{S}$, the non-additivity parameter $\Delta$, and the packing fraction $\eta$, as design parameters.
The values of these parameters are sampled from a multivariate normal distribution in each generation according to the equations of sepCMA-ES.

To specifically target the hexagonal lattice, we employ the output of the trained CNN, which represents the probability that the diffraction pattern of a given configuration is classified as a hexagonal lattice, denoted as $P_{\text{HEX}}$, as the fitness function, i.e.  $f = \bar{P}_{\text{HEX}}$. 
Here, the "bar" indicates that the input diffraction pattern for the classifier is an averaged diffraction pattern of fifty different configurations collected during the production phase of the simulation.
It is important to note that the CNN is trained using diffraction patterns of fluid, crystalline, and quasicrystalline phases obtained from a monodisperse system interacting with a hard-core square-shoulder potential.
We thus investigate whether the classification of the CNN trained with diffraction patterns of a monodisperse system is general enough to inverse design target phases in a binary system.

The results of the reverse-engineering process are presented in Fig.~\ref{fig:fitness-hex}.
The algorithm successfully finds the target hexagonal lattice in approximately eight generations, evident from the rapid rise in the mean fitness to a high value.
By the tenth generation, the algorithm consistently samples state points that, on average, facilitate the self-assembly of the hexagonal lattice.
Additionally, Fig.~\ref{fig:fitness-hex}(c) shows a representative snapshot along with its diffraction pattern computed for the large species, exhibiting  clear hexagonal symmetry.

The evolution of the design parameters is  shown in Fig.~\ref{fig:fitness-hex}(b).
The algorithm quickly converges  to the range of design parameters where the hexagonal lattice self-assembles. 
Comparing these findings to those obtained in~\textcite{fayenSelfassemblyDodecagonalOctagonal2023a} it is  expected that the inverse design protocol would  favor the self-assembly of this phase as the stability region of the hexagonal lattice  is extensive in the phase diagram.
In our work, this is evident from the wide  spread of samples with high fitness values, which gradually narrows as the evolution progresses  and the optimization method converges.
Subsequently, the probability distribution narrows around the estimated mean.
This behavior is attributed to the rugged landscape of $f$, with multiple minima, allowing the hexagonal lattice to self-assemble across a broad range of  design parameters. 
It is important to note that once the optimizer has focused on a local minimum,  other samples from other minima  get discarded due to the selection mechanism choosing the best samples within each generation.
These challenges can be overcome by using different exploration mechanisms which could explore other minima during optimization, even after discovering a local minimum of $f$.~\cite{augerRestartCMAEvolution2005,hansenBenchmarkingBIpopulationCMAES2009}  

\subsection{Inverse design of a dodecagonal quasicrystal}
\begin{figure*}
\centering
    \includegraphics[width=\textwidth]{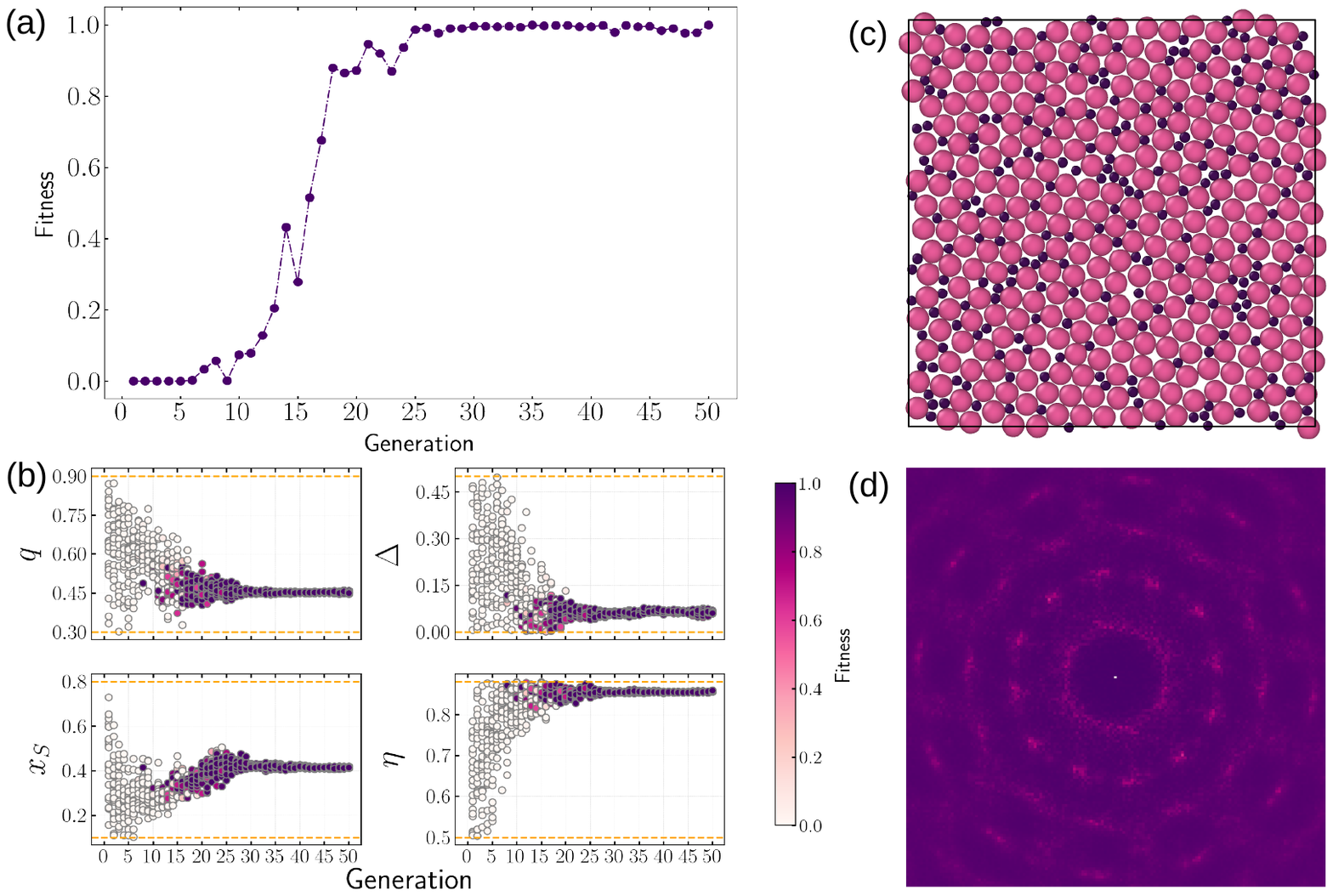}
    \caption[hex]{\textbf{Inverse design of the QC12 in a non-additive hard-disk mixture using the convolutional neural network as order parameter.}
       \begin{enumerate*}[label=(\alph*)]
         \item Evolution of the mean fitness during the inverse design protocol.
         \item Evolution of the design parameters, size ratio $q=\sigma_S/\sigma_L$, non-additivity parameter $\Delta$, small species composition $x_S$, and packing fraction $\eta$, for the inverse design of the QC12. Each dot represents a sample for a given generation. The dotted lines are the bounds imposed for the optimization problem.
         \item Exemplary configuration snapshot of a QC12 obtained during the last generation. \label{conf-qc12}
        \item Diffraction pattern of the configuration in~\ref{conf-qc12}.
        \end{enumerate*}}
    \label{fig:fitness-qc12}
\end{figure*}

We now target a dodecagonal quasicrystal (QC12) phase in a non-additive binary hard-disk mixture, recently predicted in simulations.~\cite{fayenInfinitepressurePhaseDiagram2020a,fayenSelfassemblyDodecagonalOctagonal2023a}
This quasicrystal comprises a random tiling of squares and equilateral triangles, exhibiting distinct  12-fold rotational symmetry.
This square-triangle random tiling has been  investigated extensively,~\cite{oxborrowRandomSquaretriangleTilings1993} and observed in computer simulations of monodisperse systems.~\cite{doteraMosaicTwolengthscaleQuasicrystals2014a}
Given that data for a one-component 12-fold quasicrystal (QC12) is already included in the training data of the CNN, it can   readily be used to target this phase within the binary mixture.

We define  the fitness function as $f = \bar{P}_{\text{QC12}}$, where $P_{\text{QC12}}$ represents the probability that the diffraction pattern of a given configuration is classified as a QC12 by the CNN.
Similar to the protocol used for the hexagonal lattice, the "bar" indicates that the input diffraction pattern for the classifier is an averaged diffraction pattern of fifty different configurations collected during the production phase of the simulation.

We present the results of the inverse design protocol in  Fig.~\ref{fig:fitness-qc12}.
We clearly observe that the optimization of the quasicrystal phase requires more generations to identify a local minimum than in the HEX case.
However, once found, the evolution of parameters remains stable, eventually reaching a plateau.
A clear indicator of  the inverse design protocol  approaching the target phase is  monitoring  the packing fraction $\eta$, which shows a consistent increase  as the protocol converges.
Higher packing fractions are  favored, since the QC12 and other crystalline phases  only self-assemble at high pressures and high packing fractions.
In Fig.~\ref{fig:fitness-qc12}(a), we observe that after approximately ten generations, $\eta$ starts to increase consistently toward values exceeding $\eta > 0.8$. Subsequently, the packing fraction remains close to $\eta \approx 0.84$.

The behavior observed in the other parameters, specifically $q$ and $x_{S}$, contrasts with that of the packing fraction $\eta$.
The size ratio $q$ displays extensive variations within the first ten generations, as shown by the large spread in the samples in Fig.~\ref{fig:fitness-qc12}(b).
During the first ten generations, at least 90\% of the total range of the size ratio $q$ is explored.

During the exploration phase, corresponding to the first fifteen generations of the sepCMA-ES, attempts are made to sample a wide region of the search space, until a high fitness state point is found, subsequently directing  the optimizer toward that region.
However, for both $q$ and $x_{S}$, the region explored remains quite large, covering at least 60\% of the total range, until it reaches a plateau at around generation number 30.
The extensive exploration is required since the region of self-assembly of the octagonal quasicrystal is small compared to the size of the design parameter space.
Still, the average values found for $q$ and $x_{S}$ agree well with Ref. \citenum{fayenSelfassemblyDodecagonalOctagonal2023a}.

Regarding the non-additivity parameter $\Delta$, low values are obtained, which is expected since the small particles stabilize the random square-triangle tiling.
Larger values of $\Delta$ would imply significant overlap between small and large particles, a scenario  the optimizer has steered away from within the search space.

We would like to emphasize once more that the QC12 self-assembled in the binary mixture comprises a random square-triangle tiling for the large species.
The data used to train the CNN originates from a single-component system that also exhibits a square-triangle tiling.~\cite{oxborrowRandomSquaretriangleTilings1993}
It is important to note that the current training data may not be sufficient to facilitate the inverse design of other quasicrystals observed in simulations.~\cite{doteraMosaicTwolengthscaleQuasicrystals2014a,schoberthMolecularDynamicsStudy2016}

\subsection{Inverse design of octagonal quasicrystal}
\begin{figure*}
\centering
\includegraphics[width=\textwidth]{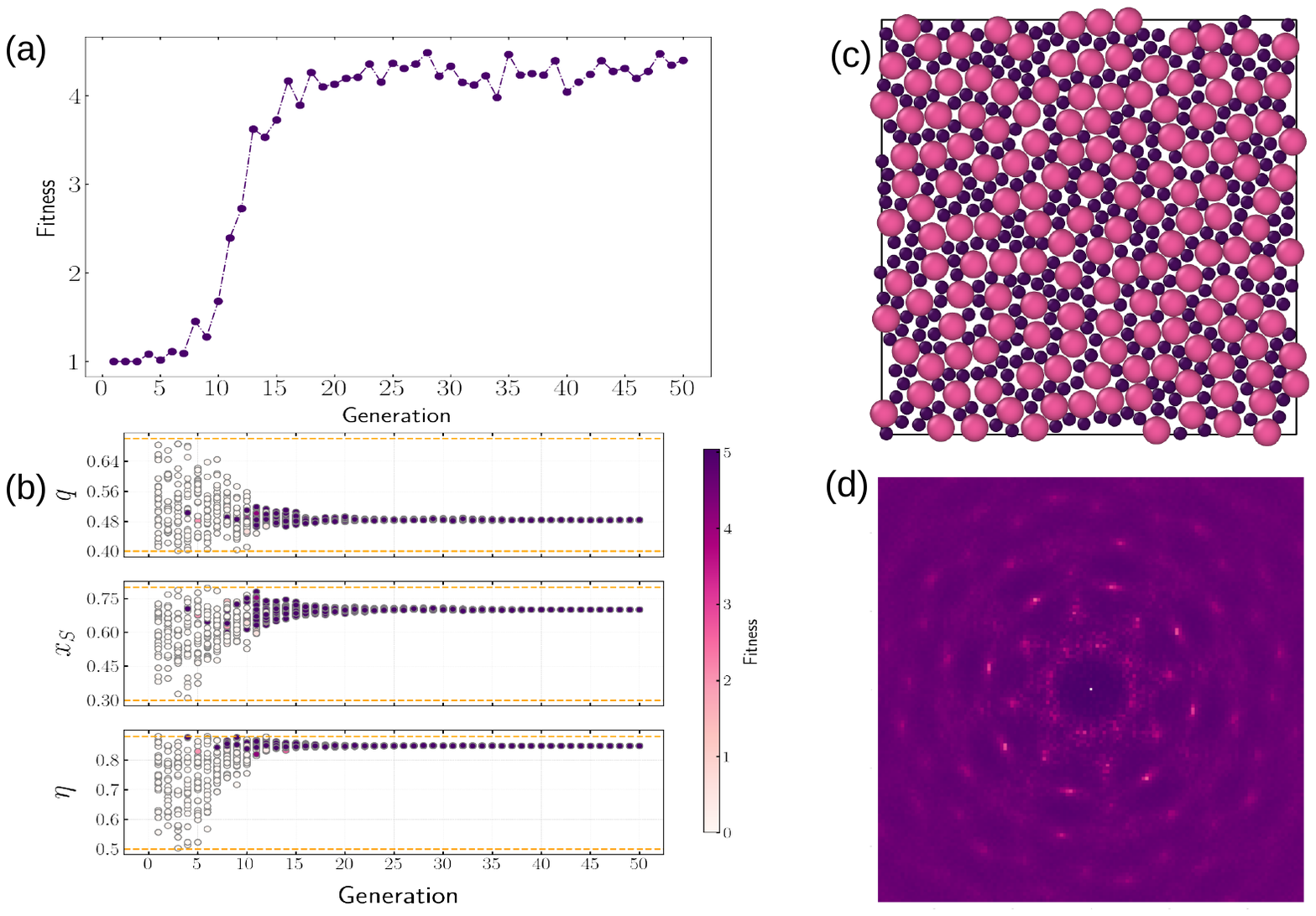}
\caption[qc8]{\textbf{Inverse design of the octagonal quasicrystal in a non-additive hard-disk mixture using the symmetry-based order parameter.}
    \begin{enumerate*}[label=(\alph*)]
        \item Evolution of the mean fitness during the inverse design protocol.
        \item Evolution of the design parameters, size ratio $q=\sigma_S/\sigma_L$, small species composition $x_S$, and packing fraction $\eta$, for the inverse design of the QC8 using the symmetry-based fitness function. Each dot represents a sample from a specific generation. The dotted lines are the bounds imposed for the optimization problem.
        \item Typical configuration snapshot of a octagonal quasicrystal obtained during the last generation. \label{conf-qc8}
        \item Diffraction pattern of the configuration in~\ref{conf-qc8}.
   \end{enumerate*}}
    \label{fig:qc8-evolution}
\end{figure*}

So far, we have successfully reverse-engineered hexagonal crystals and  dodecagonal quasicrystals within a non-additive binary hard-disk mixture using a CNN trained solely with diffraction patterns from  a single-component system. 
However, in cases where we aim to inverse design structures within a binary mixture that are not available in the one-component system, we face a challenging recursive loop.
The sought-after structure is necessary to train the CNN, while a trained CNN is crucial to inverse design  that specific structure.
An example of such a structure is the octagonal quasicrystal (QC8),~\cite{fayenSelfassemblyDodecagonalOctagonal2023a} for which diffraction patterns from a single-component system are absent in our training data.
This absence poses a challenge when attempting to employ the CNN for the inverse design of QC8 within the binary mixture.
We therefore resort to our symmetry-based order parameter, which bypasses the requirement for training a CNN and can be used as a fitness function in our inverse design protocol.

Our focus is now on achieving the octagonal quasicrystal structure using the fitness function presented in Section~\ref{sec:DSASS}.
Consistent with our previous endeavours, our objective is to identify the optimal set of design parameters that facilitates the self-assembly of QC8s.
We focus on the following design parameters, the size ratio $q$, the small species composition $x_{S}$, and the packing fraction $\eta$.
These parameters are again sampled from a multivariate Gaussian distribution at each generation.
We compute the non-additivity parameter as a function of the size ratio $q$, using the expression $\Delta(q)=(2 \sqrt{q})/(1 + q) - 1$ to facilitate a comparison with Ref.~\citenum{fayenSelfassemblyDodecagonalOctagonal2023a}.

The fitness function is now defined as $\bar{f}_{8}$, using the expression from Eq.~\eqref{eq:fitnessfunction} for $f$.
In this case, the CNN is no longer used and the symmetry-based order parameter serves as the fitness function steering the optimization.
Similar to the previous sections, the "bar" in the fitness function indicates that the input diffraction pattern for the classifier is an averaged diffraction pattern of fifty different configurations collected during the production phase of the simulation.

In Fig.~\ref{fig:qc8-evolution}(b), we present the evolution of the parameters for the QC8, displaying a plateau in the evolution of the fitness after twenty generations as shown in Fig.~\ref{fig:qc8-evolution}(a).
The packing fraction $\eta$ undergoes extensive exploration but quickly localizes at high values, as expected since the QC8 can only self-assemble at high packing fractions.
Additionally, the size ratio $q$ and small species composition $x_{S}$ converge to values with high fitness after fifteen generations.  These values align with Ref.~\citenum{fayenSelfassemblyDodecagonalOctagonal2023a}.
A representative snapshot and diffraction pattern are shown in Fig.~\ref{fig:qc8-evolution}(c) and Fig.~\ref{fig:qc8-evolution}(d), respectively, highlighting the octagonal symmetry.

Our results thus show that the fitness, defined in terms of the symmetry of the diffraction pattern, functions as intended.  
From Fig.~\ref{fig:qc8-evolution}(a), we observe that the evolution of the fitness function appears noisy in the later generations, a characteristic not observed in our results using the CNN. 
This could stem from the fitness function not being bounded but instead relying on obtaining peaks in the power spectral density $P(m)$ as large as possible. 
However, as these values can fluctuate due to penalties imposed on other frequencies, the maximum values of the peaks are not known in advance.
One way to solve this issue could involve computing multiple scores simultaneously using different sets of diffraction patterns, or  averaging the total fitness value during the evolution of the design parameters.

\section{\label{sec:conclusions} Conclusions}
In conclusion, we have extended a recently introduced inverse design protocol to reverse-engineer crystalline and quasicrystalline structures within a two-dimensional non-additive binary hard-disk mixture.
This method employs a CNN to characterize and classify diffraction patterns  as similar or dissimilar to a target phase.
This information can be used together with an optimizer, e.g. an evolutionary algorithm, to find the optimal state point and interaction parameters that facilitate the assembly of the target phase. 
More importantly, we show that we have successfully inverse designed hexagonal crystals and dodecagonal quasicrystals within a binary mixture using a CNN trained on solely diffraction patterns from a single-component system. 

When dealing with unknown phases, the neural network cannot be trained since there is no readily available information about the system.
To address this limitation, a symmetry-based order parameter was introduced, enabling the inverse design of crystalline structures and quasicrystals with new symmetries.
This order parameter determines the symmetry of the diffraction patterns by identifying the number of reflection axes.
More specifically, this approach bypasses the necessity of training a CNN and the requirement for diffraction patterns of the structures to be reverse-engineered.
By only changing the fitness function in our inverse design protocol, we successfully inverse designed octagonal quasicrystals within a binary mixture.

This study demonstrates that data from single-component systems is sufficient for training neural networks and reverse engineering structures in binary mixtures, provided that these phases are represented in the training data and large quantities can be obtained. 
While the neural network performs well when data of the specific structure is available, the symmetry-based order parameter serves as a valuable tool when the target structure data remains elusive.


\begin{acknowledgments}
M.D., A.P.A.O, and E.A.B.M acknowledge funding from the European Research Council (ERC) under the European Union’s Horizon 2020 research and innovation programme (Grant agreement No. ERC-2019-ADG 884902 SoftML).
A.P.A.O, and E.A.B.M thank SURF (\url{www.surf.nl}) for the support in using the National Supercomputer Snellius.
\end{acknowledgments}

\section*{Data Availability}
The data that support the findings of this study are available
from the corresponding author upon reasonable request.

\bibliography{references}

\end{document}